\documentclass{WileyMSP-template}
\usepackage[version=3]{mhchem} % Formula subscripts using \ce{}
\usepackage[bottom]{footmisc}

\begin{document}

%\pagestyle{fancy}
%\rhead{\includegraphics[width=2.5cm]{vch-logo.png}}

\title{Growth and helicity of non-centrosymmetric Cu$_2$OSeO$_3$ crystals}

\maketitle

\author{Aisha Aqeel, Jan Sahliger, Guowei Li, Jacob Baas, Graeme R. Blake, Thomas. T.M. Palstra and Christian H. Back}

\dedication{}

\begin{affiliations}
Dr. A. Aqeel, J. Sahliger\\
Physik-Department, Technische Universit\"at M\"unchen, D-85748 Garching, Germany\\

Dr. G. Li\\
Max Planck Institute for Chemical Physics of Solids, 01187 Dresden, Germany\\
ing. J. Baas, Dr. G. R. Blake, Prof. Dr. Thomas. T.M. Palstra \thanks{present address: University of Twente, 7522NB Enschede, The Netherlands }\\
Zernike Institute for Advanced Materials, University of Groningen, Nijenborgh 4, 9747 AG Groningen, The Netherlands\\
$^1$ Present address: University of Twente, 7522NB Enschede, The Netherlands 

Prof. Dr. C. H. Back\\
Physik-Department, Technische Universit\"at M\"unchen, D-85748 Garching, Germany\\
Munich Center for Quantum Science and Technology (MCQST), D-80799 M\"unchen, Germany\\
Email Address: christian.back@tum.de\\
\end{affiliations}


\begin{abstract}
We have grown Cu$_2$OSeO$_3$ single crystals with an optimized chemical vapor transport technique by using SeCl$_4$ as a transport agent. Our optimized growth method allows to selectively produce large high quality single crystals. The method is shown to consistently produce Cu$_2$OSeO$_3$ crystals of maximum size 8~mm~x~7~mm~x~4~mm with a transport duration of around three weeks. We found this method, with SeCl$_4$ as transport agent, more efficient and simple compared to the commonly used growth techniques reported in literature with HCl gas as transport agent. The Cu$_2$OSeO$_3$ crystals have very high quality and the absolute structure are fully determined by simple single crystal x-ray diffraction. We observed both type of crystals with left- and right-handed chiralities. Our magnetization and ferromagnetic resonance data show the same magnetic phase diagram as reported earlier~\cite{Qian2016}. 

\end{abstract}
\section{Introduction}
Investigation of complex magnetic systems is generally limited by an inability to obtain sufficiently large, pure high quality crystals. This is especially true for the noncentrosymmetric magnets with chirality. In this class of materials the interactions that may lead to symmetry breaking magnetic order do not cancel each other when evaluated over the unit cell. The most well studied chiral systems are MnSi\cite{Muhlbauer2009,Jonietz2010}, Mn$_{1-x}$Fe$_x$Ge \cite{Shibata2013}, FeGe \cite{Yu2011} and semiconducting Fe$_{1-x}$Co$_x$Si~\cite{Munzer2010}. In these chiral magnets, the principal magnetic phases are a helical phase, a single domain conical phase and a skyrmion state (known as A-phase). They appear in a small magnetic field-temperature (B-T) pocket close to transition temperature $T_{\rm c}$. In the chiral atomic framework of this crystal family, the orbital motions of localized electrons also take helical paths.
The neighbouring spins of localized electrons are coupled by the relativistic spin-orbit interaction called Dzyaloshinskii-Moriya (DM) interaction \cite{Dzyaloshinsky1958,Moriya1960}. As the sign of the DM interaction is determined by the chemical composition, it emphasizes that the magnetic chirality is intrinsically dependent on the lattice handedness. It has been shown experimentally in Mn$_{1-x}$Fe$_x$Ge crystals that the skyrmion helicity is directly determined by the crystal helicity \cite{Shibata2013}.\\

Cu$_2$OSeO$_3$ is one of the most important members of the chiral group with the P2$_1$3 chiral cubic crystal structure. It is the first insulator in which the skyrmion lattice has been observed \cite{Adams2012,Seki2012} with a very similar B-T phase diagram as the other related members of this chiral group. Recently, some new magnetic phases like tilted conical spiral~\cite{Qianeaat7323}, low-temperature skyrmion lattice phase~\cite{Chacon2018} and elongated skyrmions~\cite{Aqeel2021FMR} have been observed in Cu$_2$OSeO$_3$. The insulating behavior of this magnetic material makes the study of the decisive role of crystal helicity especially more interesting by excluding other contributions due to conduction electrons. To understand the unique magnetic structure of Cu$_2$OSeO$_3$, several different techniques have been employed including, mu-SR~\cite{Maisuradze2011}, Lorentz transmission electron microscopy~\cite{Seki2012}, ac-susceptibility measurements~\cite{Levatic2014}, Terahertz Electron Spin Resonance~\cite{Ozerov2014}, time-resolved magneto-optics~\cite{ogawa2015}. Recently, generation of spin currents have been studied in Cu$_2$OSeO$_3$ by spin pumping experiments~\cite{Hirobe2015}.

Considering the large interest in the magnetic properties of Cu$_2$OSeO$_3$, it is important to look for new, efficient and fast single crystal growth techniques. Conventionally, Cu$_2$OSeO$_3$ crystals are grown by vapor transport method with HCl gas as transport agent. With this growth method only one helicity has been reported \cite{Dyadkin2014}.  The other helicity has not been reported to the best of our knowledge. It is known that the structural and magnetic chiralities for Cu$_2$OSeO$_3$ crystals are directly related with each other \cite{Dyadkin2014}. Therefore to use both magnetic chiralities, it is needed to improve the growth techniques not only to speed up the growth rate but also to get crystals with both chiralities.
%To control the magnetic helicity , it is important to find a growth method to tune the crystal chirality from left- to right-handedness.
Here, we report a new and fast way for the growth of Cu$_2$OSeO$_3$ single crystals with SeCl$_4$ as transport agent. We observed very high quality crystal growth yielding both chiralities with this new growth technique. The crystal structure of Cu$_2$OSeO$_3$ crystals has been studied before \cite{Meunier1976,Effenberger1986,Bos2008} with different diffraction techniques. Here, we used the simplex single crystal x-ray diffraction (XRD) to establish the absolute structures for both handedness, which also emphasizes good quality of the crystals.

\section{Experimental}
Single crystals of Cu$_2$OSeO$_3$ were grown by the standard chemical-vapor transport method. However, the novelty of this growth is the use of selenium tetrachloride (SeCl$_4$) as a transport agent. Previously, SeCl$_4$ is mainly used to grow the  molybdenum and tungsten diselenides. In literature Cu$_2$OSeO$_3$ is usually grown by HCl gas \cite{Belesi2010}, here, we report the growth of chiral magnets with SeCl$_4$ as transport agent which is new and different from literature \cite{Legma1993}. For growth, transparent quartz ampoules (30~mm inside diameter, 30~cm length)  were used. They were first carefully cleaned with ethanol, acetone, 10$\%$ HF and demi water and dried overnight at 200 $^o$C before the charge was introduced.
SeCl$_4$ is very hygroscopic; therefore, it was weighed and introduced into the transport tubes in a glove box under a nitrogen atmosphere. Mixtures of high-purity CuO (Alfa-Aldrich, 99.995 $\%$) and \ce{SeO2} (Alfa-Aldrich, 99.999$\%$) powders in a molar ratio of 2:1 were sealed in an evacuated quartz ampoule with 0.54~g of SeCl$_4$ (Alfa-Aldrich, 99.5$\%$). After a few minutes of degassing, the part of the ampoule containing the chemicals was immersed into liquid nitrogen, subsequently evacuated and sealed after the chemicals had cooled below evaporation temperatures. The ampoule was then placed horizontally into a tubular three-zone furnace having 18~cm~-~long zones separated by a distance of 3~cm. The temperature of the furnace was raised gradually by 50 $^o$C/h to 600 $^o$C. To get rid of unwanted nucleation centers, a reverse temperature gradient was applied by adjusting the temperature of the source zone (T$_\text{hot}$) to 610 $^o$C and the deposition zone (T$_\text{cold}$) to 660 $^o$C for 24h. Afterwards, T$_\text{hot}$ and T$_\text{cold}$ were adjusted to 610 $^o$C and 570 $^o$C, respectively, for growth.  These furnaces were regulated by a PID electronic regulator (SHINKO) with $\pm$0.5 $^o$C temperature stability at 500-650 $^o$C. After two weeks, shiny crystals could be seen at the deposition zone. After four weeks, the ampoules were quenched at the source zone so that all gas vapours would quickly condense at the source zone.
\begin{table}[htbp]
  \caption{Growth conditions for Cu$_2$OSeO$_3$ with CVT growth method for different transport agents (TA).}
  \label{tbl}
  \begin{tabular}{lllll}
    \hline
    TA  & T$_\textrm{hot}$ ($^o$C) & T$_\textrm{cold}$ ($^o$C) &  duration of growth (d) & max. size of crystals (mm$^3$)\\
    \hline
    HCl \cite{Belesi2010}   & 620   & 580 & 49 & 130-150 \\
    SeCl$_4$ & 610 & 570 & 23 & 210-224\\
    \hline
  \end{tabular}
\end{table}
The extreme hygroscopic nature of SeCl$_4$ resulted in the presence of water in the ampoules, in spite of all precautions taken. The presence of water can create the vapor phase of hydrogen chloride (HCl) and a chalcogen oxichloride (SeOCl), thus making the analysis of the transport mechanism more complex. However, we observed that the presence of moisture slows down the transport process. This transport method with  SeCl$_4$ resulted in reasonably big and thick crystals.  To compare the efficiency of the growth method, we also synthesized the crystals with HCl gas as transport agent as reported in literature  \cite{Belesi2010}. The growth conditions are summarized in Table~\ref{tbl}. The crystal structure of Cu$_2$OSeO$_3$ crystals is investigated by D8 Venture single crystal x-ray diffraction (XRD). The crystal quality is checked with precision scans of XRD for full sphere approximation. The morphology and elemental analysis were examined using a Philips~XL~30 scanning electron microscopy (SEM) equipped with a EDS system, which was operated at an accelerating voltage of 20~kV. The magnetization measurements were done using a Quantum Design MPMS-XL~7 SQUID magnetometer.

To further study the phase magnetic phase diagram of Cu$_2$OSeO$_3$, we performed ferromagnetic
resonance (FMR) measurements using a broadband spin-wave spectroscopy technique~\cite{Schwarze2015,Aqeel2021FMR} on both HCl and SeCl$_4$ grown crystals. For this purpose we used two polished crystals with similar dimensions: HCl grown sample with size - 1.5 $\times$ 2.5 $\times$ 0.5 mm$^3$ and SeCl$_4$ grown crystal with size - 2.9 $\times$ 2.7 $\times$ 1 mm$^3$. The crystals were polished using the technique reported in Ref.~\cite{Aqeel2014}. The coplanar waveguides with a signal line of 50~$\mu$m width and gap of 25~$\mu$m width were patterned. The CPWs were directly patterned onto the oriented polished crystals with (110) and (111) surface, respectively. They were patterned by e-beam lithography followed by e-beam evaporation of Ti (10nm)/Au (150nm). The excitation field distribution of the CPWs is shown in Fig.~\ref{fig:FMR}(b). The samples with CPW were mounted on a continuous flow cryostat. A vector network analyzer (VNA) was used to measure the resonance signals. The temperature reading of the cryostat was different from the MPMS system used to measure the magnetization data due to the placement of the temperature sensor. The temperature difference (7K -10K) between both setups is adjusted in the manuscript. A rotatable electromagnet was used to provide the static magnetic field up to 500~mT. 
\section{Results}
\begin{figure}
\includegraphics*[width=0.7\columnwidth]{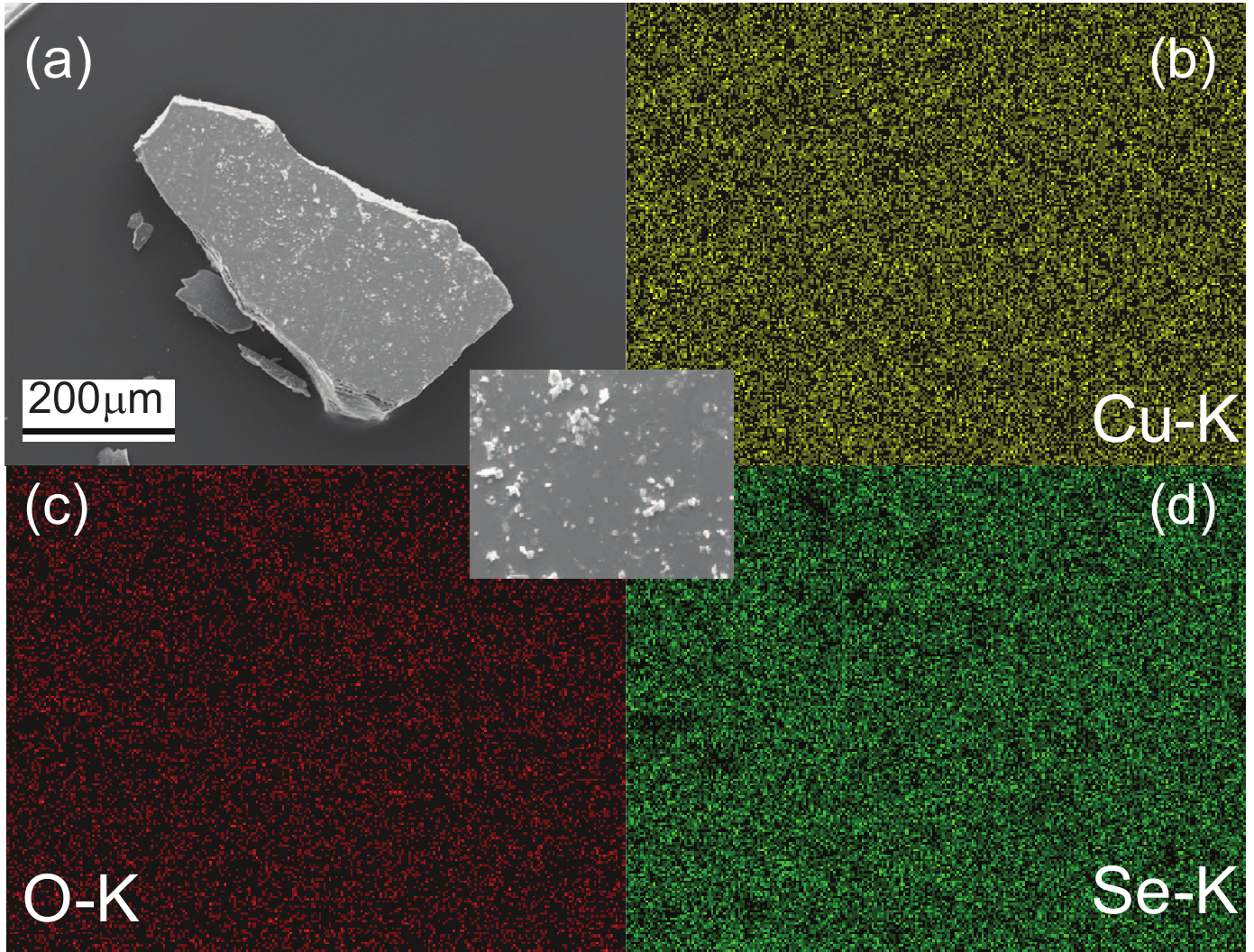}
\caption{\label{fig:6}
(a) The SEM image of a typical single crystal. EDS elemental mapping for Cu (b), O (c), and Se (d) elements in the as synthesized crystal. The inset show the scanning area for element mapping.}
\end{figure}

A tiny single crystal grown with SeCl$_4$ was selected for morphology and element analysis as shown in Figure~\ref{fig:6}(a). The as synthesized crystal has a rough surface with many tiny nanoparticles attached on it. The molar ratio of Se, Cu, and O are determined to be very close to the stoichiometric Cu$_2$OSeO$_3$ sample. Furthermore, the energy dispersive spectrometer(EDS) elemental mapping (Figure~\ref{fig:6}(b)-(d)) demonstrates that the Cu, O, and Se atoms are uniformly distributed, which unambiguously reveals the uniformity of the single crystal. That is, a homogeneous and high-quality sample was successfully synthesized with such a simple method.
Table \ref{tb2} shows the parameters used to establish the absolute structure of Cu$_2$OSeO$_3$ single crystals.  Cu$_2$OSeO$_3$ crystals display the P2$_1$3 space group and the ions occupy the Wyckoff positions that are summarised in Table \ref{tb3}. The precision scans of XRD for full sphere approximation shows the high quality of  Cu$_2$OSeO$_3$ single crystals without any twinning.
\begin{table}[htbp]
  \caption{Crystallographic data and structure refinement for Cu$_2$OSeO$_3$ single crystals.}
  \label{tb2}
  \begin{tabular}{ll}
  \hline
  \hline
    Temperature & 100 K \\
    crystal system & cubic \\
    space group & $P2_13$ \\
    wave length & 0.7107 $A^o$\\
    unit cell dimension a & 8.9446 $A^o$\\
    $\theta$ range for data collection & 3.147$^o$ - 32.25$^o$ \\
    Limiting indices & -13 $\leq$ h $\leq$ 13\\
    &-11 $\leq$ k $\leq$ 11\\
    &-13 $\leq$ l $\leq$ 13\\
    Reflections collected~/~unique & 0.0367 \\
    Final R indices [I $>$ 2$\sigma$(I)] & 0.0312\\
    Absolute structure parameter & -0.01(2)\\
    \hline
    \hline
 \end{tabular}
\end{table}

The chirality of the crystals has been characterized by the Flack parameter analysis. The Flack parameter is defined as the ratio between two opposite-handed domains  for non-centrosymmetric crystals giving rise to a resonant contribution in the x-ray scattering amplitudes. A Flack parameter equal to zero corresponds to a single domain of the chiral structure (enantiopure) and a Flack parameter equal to 1 represents a single domain structure but with opposite chiralities. The absolute structures are solved by calculating the atomic coordinates during refinement of the Flack parameter x by using the twin model for intensities of hkl reflections as follows:
\begin{equation}
\textrm{I}_\textrm{hkl}^\textrm{calc}=(\textrm{1-x})\mid \textrm{F}_\textrm{hkl} \mid ^2 + \textrm{x} \mid \textrm{F}_\textrm{-h-k-l} \mid^2  \label{eq1}
\end{equation}
Here, $\mid \textrm{F}_\textrm{hkl} \mid$ and $\mid \textrm{F}_\textrm{-h-k-l} \mid$ represent the structure factors. The dual-space SHELXL method was used for the structure determination.  The full sphere of Bragg reflections was used for refinement. Results of least square refinement gives a Flack x of 0.013(17) indicating two absolute structures having opposite chirality. The deviation factor is defined as:
\begin{equation}
R_1 = \Sigma \mid \textrm{F}_\textrm{obs}-\textrm{F}_\textrm{calc} \mid / \Sigma \textrm{F}_\textrm{obs}  \label{eq1}
\end{equation}
The standard deviation $R_1$ was found to be 0.0217, which shows that the scattering strictly follows  the Flack conditions. We measured eight crystals to resolve the absolute structure, in which we found five right-handed and three left-handed enantiomers. The atomic coordinates for absolute structures for left-handed and right-handed enantiomers of Cu$_2$OSeO$_3$ are summarized in Table~\ref{tb3}.

\begin{table}[htbp]
  \caption{Atomic coordinates and Wyckoff positions (WF) for Cu$_2$OSeO$_3$ for both handedness.}
  \label{tb3}
  \begin{tabular}{ll | lll | lll}
  \hline
  \hline
  & & & Right-handed & & &  Left handed &\\
 & WP & x & y & z & x & y & z \\
   \hline
Cu (1) & 4a &0.88589(3) & 0.88589(3) & 0.88589(3) &    0.11404(4)& 0.11404(4) & 0.11404(4) \\
Cu (2) & 12b & 0.13439(3) & 0.12108(3) &  0.87247(3) &  0.86549(4) & 0.87895(4) & 0.12754(4) \\
Se (1) & 4a & 0.45963(3) & 0.45963(3) & 0.45963(3) &   0.54031(4) & 0.54031(4) & 0.54031(4)\\
Se (2) &  4a & 0.21201(3) & 0.21201(3) & 0.21201(3) &  0.78802(4) &0.78802(4) & 0.78802(4) \\
O (1) & 4a & 0.01031(3) & 0.01031(3)  & 0.01031(3) &   0.98974(3) & 0.98974(4)& 0.98974(3)\\
O (2) & 12b & 0.76232(2) & 0.76232(2) & 0.76232(2) &    0.23730(3) &  0.23730(4) &  0.23730(3) \\
O (3) & 4a & 0.27029(2) & 0.48318(2)  & 0.46954(2) &    0.72971(3) & 0.51663(4) & 0.53014(3)\\
O (4) & 12b & 0.27257(2) & 0.18681(2) & 0.03276(2) &   0.72786(3) &  0.81329(4) & 0.96738(3)\\
    \hline
    \hline
 \end{tabular}
\end{table}
\begin{figure}
\includegraphics*[width=0.5\columnwidth]{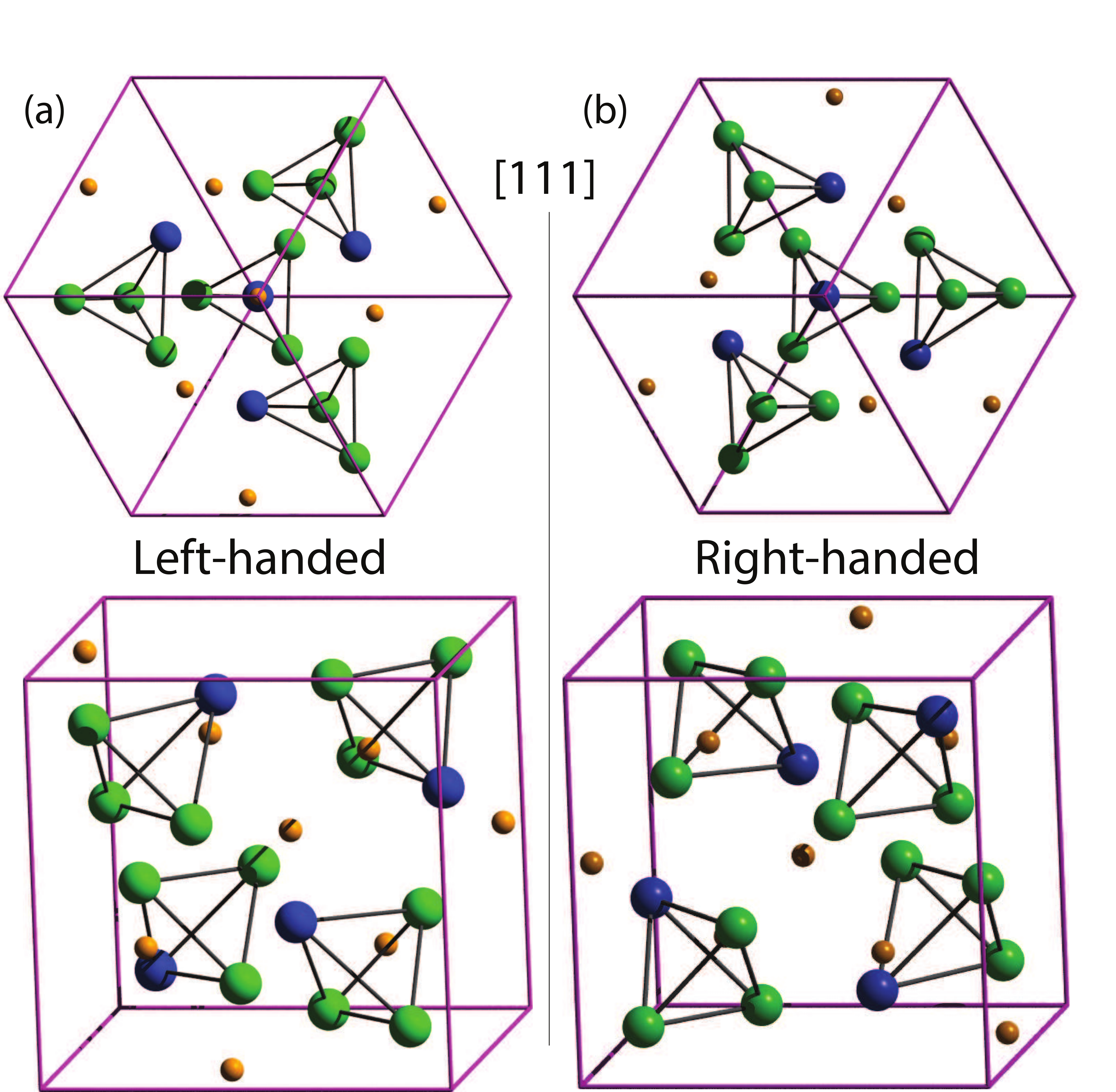}
\caption{\label{fig:1}
The two chiral crystal structures of Cu$_2$OSeO$_3$ where green and blue spheres represent Cu1 and Cu2 atoms. The top views are along the body diagonal of the cube (along [111] axis). (a) right-handed and (b) left-handed crystals.}
\end{figure}

\begin{figure}[hpb]
\includegraphics*[
width=0.98\columnwidth,
trim = 0mm 1mm 0mm 0mm, clip]{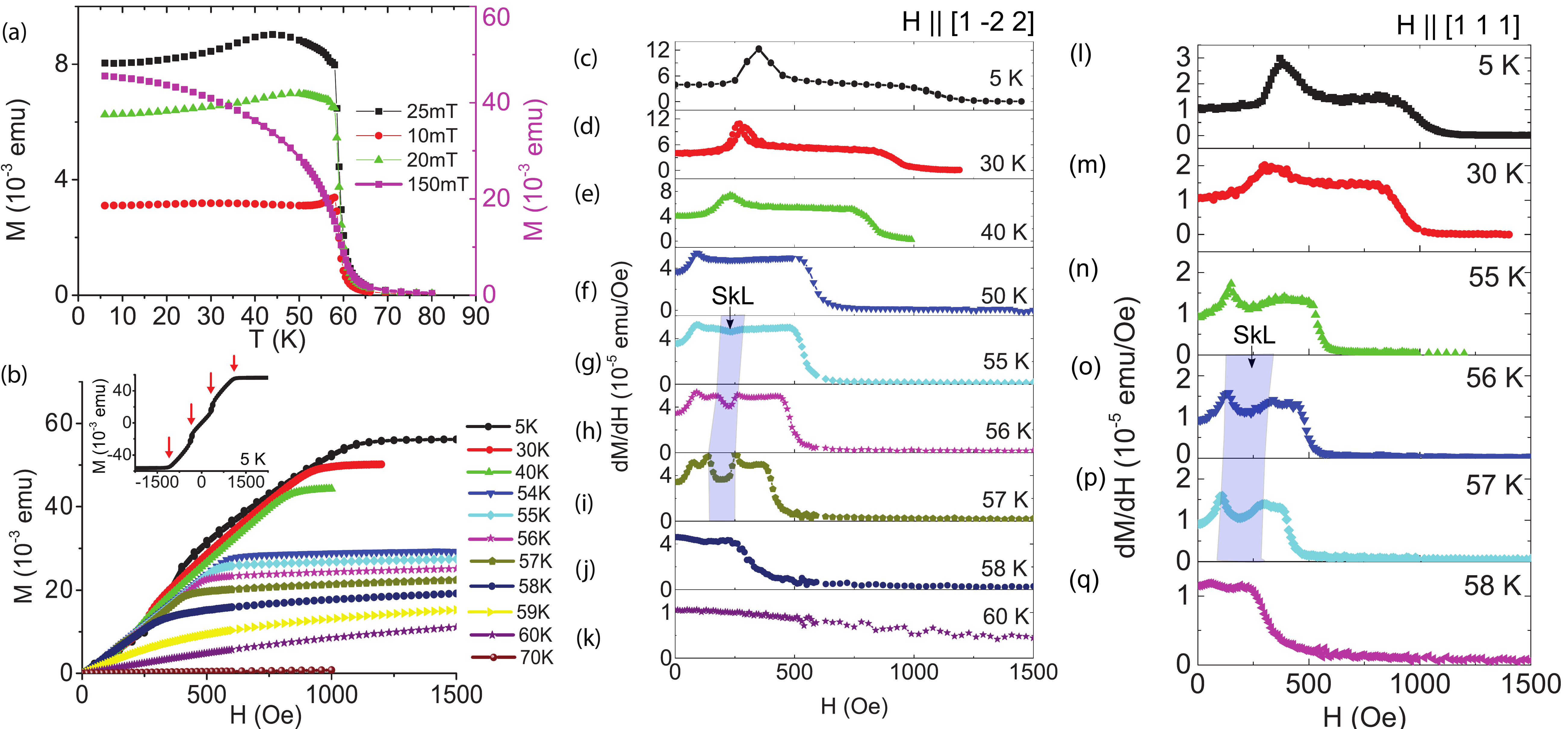}
\caption{\label{fig:3}
(a) Temperature dependences of the FC magnetic susceptibilities in different applied magnetic fields along [001] direction. (b) Magnetic field dependence of the  magnetization at different temperatures, with magnetic field applied along [1-22] direction. The inset shows the slope-change behavior at the fields $\sim$350~Oe and 1100~Oe at 5~K, as indicated by arrows. The dM/dH vs H at different temperatures with magnetic field applied along (c)-(k) [1-22]  and  (l-q) [111] crystallographic directions. The shaded region represents the skyrmion lattice phase.  }
\end{figure}
% \begin{figure}[hpb]
% \includegraphics*[
% width=0.7\columnwidth,
% trim = 0mm 1mm 0mm 0mm, clip]{2.eps}
% \caption{\label{fig:2}
% (a) Temperature dependences of the FC and ZFC magnetic susceptibilities in 10~mT applied magnetic field along [111] direction. (b) Magnetic field dependence of the  magnetization at different temperatures with magnetic field applied along [111] direction. (c)-(k) The dM/dH vs H at different temperatures. }
% \end{figure}
Figure~\ref{fig:3}(a) shows the temperature dependence of field-cooled magnetization measurements under an applied field H varying from 100 Oe to 1500~Oe with a \(T_{c} \approx 60~K\).

% Figure~\ref{fig:2}(a) shows no difference between field-cooled (FC) and zero-field-cooled (ZFC) magnetization curves measured at 10~mT.
% ZFC measurements are done by cooling the sample in zero field and applying the magnetic field while measuring by warming up. Whereas FC measurements were performed by cooling and measuring in same applied magnetic fields.

\begin{figure}[hpb]
\includegraphics*[
width=0.98\columnwidth,
trim = 0mm 1mm 0mm 0mm, clip]{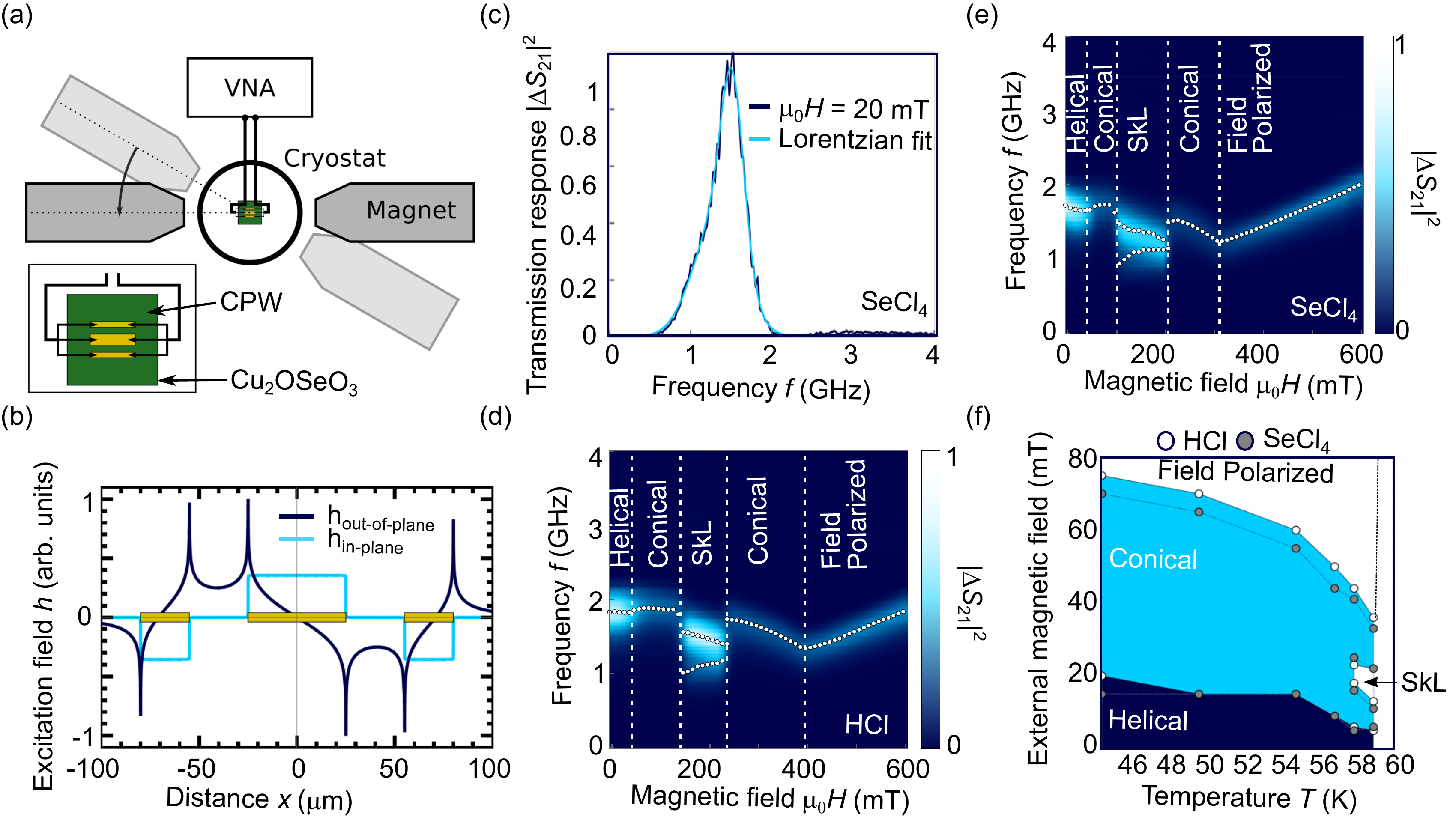}
\caption{\label{fig:FMR}
Schematics of the broad band ferromagnetic resonance setup with a vector network analyzer (VNA), low temperature measurement setup (cryostat) and a rotatable electromagnet. The inset shows the schematics of the coplanar waveguide (CPW) structured directly onto the sample surface. The static magnetic field H is applied along the CPW within the plane of the sample along the [1-11] crystallographic direction of Cu$_2$OSeO$_3$. (b) Field distribution of the CPW. (c) Transmission difference $\Delta |S_{21}|^2$ as a function of frequency measured at 20mT at 5 K. (d-e) Color-coded resonance map obtained from the line scans of the transmission response measured as a function of frequency at 58 K for the HCl and SeCl$_4$ grown samples. (f) The magnetic phase diagram of Cu$_2$OSeO$_3$ single crystals obtained by applying static magnetic field along [1-11] crystallographic direction of Cu$_2$OSeO$_3$. The phase boundaries (dots) are extracted from the resonance spectra shown in~\ref{fig:FMR}(d,e). }
\end{figure}

The FMR spectra were measured as reported earlier~\cite{Aqeel2021FMR} and the background free transmission signal was defined as \(| \Delta S_{21} |^2=(|S_{21}(H)-|S_{21}(H_0)|)^2\). S$_{21}(H)$ and $S_{21}(H_0)$ are the complex transmissions measured with VNA at a fixed magnetic field H and H0, respectively. An example of such spectra measured for HCl and SeCl$_4$ grown samples are shown in Figs.~\ref{fig:FMR}(d) and\ref{fig:FMR}(e), respectively. These spectra are used to construct the high temperature magnetic phase diagram of Cu$_2$OSeO$_3$ shown in Fig~\ref{fig:FMR}(f).  

\section{Discussion}
The vapor transport technique~\cite{Belesi2010} commonly used for the growth of Cu$_2$OSeO$_3$ single crystals is relatively slow and complex due to use of the HCl gas as transport agent. However, the method reported in this paper is very simple and easy due to use of solid transport agent SeCl$_4$ and also found to be relatively fast. A disadvantage of SeCl$_4$ transport agent could be the strong silica attack and its strong hygroscopic nature which can be easily settled by using the transport agent in an inert and dry atmosphere. SeCl$_4$ is frequently used in the past as an efficient transport agent for the growth of diselenides as WSe$_2$ and MoSe$_2$~\cite{Klein1998,Legma1993,Prasad1988}. Like SeCl$_4$, TeCl$_4$ can also be an efficient transport agent.  TeCl$_4$ is more stable and less hygroscopic compared to SeCl$_4$ which makes it more suitable transport agent compared to SeCl$_4$ for vapor transport growth. However, TeCl$_4$ can dope the  crystals and therefore, SeCl$_4$ is more suitable for growth of un-doped Cu$_2$OSeO$_3$ crystals. The decomposition of SeCl$_4$ will give a mixture of Selenium and dichlorine that can result in possible gaseous oxygen compounds during transport for SeCl$_4$ can be SeO$_2$, SeOCl$_2$ and SeO. Chlorine resulting from the decomposition of SeCl$_4$ is probably playing the efficient role in the transport but the role of the Selenium is not very clear in the transport.  In case of presence of water, the transport would be more complicated by also involving HCl vapors for the transport. We observed a clear decrease in the deposition rate by exposing the SeCl$_4$ transport agent to the air.

The absolute structures were solved for six different crystals, grown with SeCl$_4$ as transport agent. During refinement the Goodness of Fit (GooF) was found to be 0.9-1.03 and the scale factor K is 0.95-1.0 and the standard deviation R1 is found to be 0.2-0.5, which confirms the high quality of these crystals. Four out of six analyzed  crystals showed the same helicity and the other two crystals showed the opposite. The helicity can be defined from the Wyckoff position of magnetic ions. In the case of Cu$_2$OSeO$_3$, Cu(1) and Cu(2) ions are located at 4a and 12b Wyckoff position as shown in Table~\ref{tb3}.
The 4a Wyckoff position of Cu(1) in Cu$_2$OSeO$_3$ is \textrm{(x, x, x/0.5+x, 0.5-x, -x/-x, 0.5+x, 0.5-x/0.5-x, -x, 0.5+x)}
where x $\approx$~0.136 or x $\approx$~1~-~0.136~=~0.863 corresponding to two enantiomers. The crystals having Cu(1) at x~=~0.863 are defined as right-handed enantiomer and others with x~=~0.136 as left-handed enantiomer as shown in Table~\ref{tb3}. The structure of Cu$_2$OSeO$_3$ with the same set of coordinates for the right-handed crystals shown in  Table~\ref{tb3} is also defined as right-handed in Ref.~\cite{Dyadkin2014}. There, the crystals are defined as right-handed on the basis of similarity of 4a Wyckoff position of Cu(1) ion in Cu$_2$OSeO$_3$ and Mn in MnSi (right -handed).

The crystal helicity can also be defined by considering the closeness of the structural symmetry of the P2$_1$3 space group  with the absolute structure of P4$_1$32 as proposed by Ref.~\cite{Chizhikov2015}. P4$_1$32 space group contains only right-handed screw axes 4$_1$, therefore the right-handed crystals of P2$_1$3 space group can be easily distinguished by comparison. The same approach is also mentioned for B20 structures~\cite{Dmitriev2012}. The set of coordinates determined with this definition for right-handed crystals is found to be consistent with the obtained absolute structure for the right-handed crystals as shown in Table~\ref{tb3}.

The magnetization and resonance data shown in Figs.~\ref{fig:3} and~\ref{fig:FMR} show the presence of the skyrmion lattice phase along with other magnetic phases of Cu$_2$OSeO$_3$. To determine the linwidths and peak positions from the transmission signal, Lorentzian peak fitting was used. An example of such peak fitting is shown in Fig~\ref{fig:FMR}(c). The resonance spectra measured for both samples grown with HCl and SeCl$_4$ qualitatively show no clear difference (cf. Figs.~\ref{fig:FMR}(d) and ~\ref{fig:FMR}(e)).  For both samples, resonance signal with similar linewidth is observed, confirming that the crystals grown with the new method have the same damping as for those grown with the HCl~\cite{Stasinopoulos2017}. The phase boundaries of the skyrmion lattice for both samples grown with HCl and SeCl$_4$ coinside with each other (see ~\ref{fig:FMR}(f)).  
\section{Conclusion}
We have demonstrated a simple route that allows the growth of Cu$_2$OSeO$_3$ single crystals in a relatively short duration.  The XRD analysis shows high quality of single crystals. We observed both right-handed and left-handed  enantiomers of  Cu$_2$OSeO$_3$ and the absolute structure was fully determined by the Flack parameter analysis of the refined XRD pattern. The growth of crystals with both left- and right-handed structural chiralities can be useful to understand the coupling between structural and magnetic chiralities. The understanding of the coupling is important to control the magnetic textures such as skyrmions  for spintronics applications.\\
% Acknowledgements
\medskip
\textbf{Acknowledgements} \par %delete if not applicable))
We thank H. Berger, T. Nilges and N. Tombros for fruitful discussions. F. Reiter for assistance with the experiments. This work has been funded by the Deutsche Forschungsgemeinschaft (DFG, German Research Foundation) under TRR80 (From Electronic Correlations to Functionality, Project No.\ 107745057, Project G9) and the excellence cluster MCQST under Germany's Excellence Strategy EXC-2111 (Project No.\ 390814868). 

%\bibliographystyle{MSP}
%\bibliography{main}

\end{document}